# A review on Epileptic Seizure Detection using Machine Learning

**Muhammad Shoaib Farooq, Aimen Zulfiqar, Shamyla Riaz**
Department of Computer Science, University of Management and Technology, Lahore, 54000, Pakistan
Corresponding author: Muhammad Shoaib Farooq (e-mail: sahib.farooq@umt.edu.pk).

**ABSTRACT** Epilepsy is a neurological brain disorder which life threatening and gives rise to recurrent seizures that are unprovoked. It occurs due to the abnormal chemical changes in our brain. Over the course of many years, studies have been conducted to support automatic diagnosis of epileptic seizures for the ease of clinicians. For that, several studies entail the use of machine learning methods for the early prediction of epileptic seizures. Mainly, feature extraction methods have been used to extract the right features from the EEG data generated by the EEG machine and then various machine learning classifiers are used for the classification process. This study provides a systematic literature review of feature selection process as well as the classification performance. This study was limited to the finding of most used feature extraction methods and the classifiers used for accurate classification of normal to epileptic seizures. The existing literature was examined from well-known repositories such as MPDI, IEEEXplore, Wiley, Elsevier, ACM, Springerlink and others. Furthermore, a taxonomy was created that recapitulates the state-of-the-art used solutions for this problem. We also studied the nature of different benchmark and unbiased datasets and gave a rigorous analysis of the working of classifiers. Finally, we concluded the research by presenting the gaps, challenges and opportunities which can further help researchers in prediction of epileptic seizure.

## I. INTRODUCTION

Epilepsy is a widespread neurological disorder that is common yet deadly if goes untreated. This disorder effects people of all ages. A complex chemical change appears in the nerve cells of the brain for a seizure to occur. These chemical changes take place in the nerve cells that are made up of positive ions and negative ions which generate electrical signals [2]. These abrupt changes lead from mild jerks to severe, generalized and prolonged convulsions. This neurological condition not only causes issues in the movement but also disturbs the control of bowel or bladder function, effects the consciousness and also disturbs the cognitive functions [7]. The population that is affected by epilepsy or epileptic seizures are seventy percent adults where thirty percent are children. The cause of epilepsy in adults and children in 70% of the cases is unknown. Clinical terminology states that if seizures are recurrent, they are termed as epilepsy. An unpredicted seizure can be a cause of many problems in daily life such as a severe head injury, falling down or a road accident because the timing of an epileptic seizure is random. It is stated that there are two classifications of seizures. One is partial [1] and other is generalized seizure. In partial seizure, also called focal seizure, only a certain portion of the brain is damaged whereas, in generalized seizure, the whole brain is damaged. Mainly, epilepsy has four important stages which include Interictal, preictal, ictal and postictal. The seizure is in ictal stage when it occurs. 15 minutes before the seizure starts is preictal stage and after the seizure ends is postictal stage. Interictal stage is the time interval between two seizures.

The most stereotyped way of detecting abnormal seizures is by manually using a device called electroencephalogram (EEG). EEG is performed by placing electrode on the scalp in a way that is either invasive or non-invasive [2]. About 50 million people are alive in the world that are living and suffering from epilepsy worldwide and about 2.4 million each year are the people who are diagnosed with epilepsy according to the World Health Organisation (WHO). It was stated in a study that mostly antiepileptic drugs or surgeries are done for the treatment of epilepsy in developing countries [3].

In [1], it is acknowledged that epileptic seizure detection techniques differ under different conditions of the dataset across the electroencephalogram (EEG). Mainly, it satisfies the fact that there are different characteristics of EEG under different conditions. It is also very important to extract the right features for better prediction of epileptic seizure. These techniques help with accurate classification of data, hence leading to better accuracy of results.

There has been great focus on EEG analysis for feature extraction that has a very important role for the detection and prediction of various brain disorders [4]. Most of the researches are carried out in differentiating normal seizures from epileptic seizures. Hypothetical testing for feature refinement of feature selection method is used and wavelet transform for the improvement of classification. The computational complexity of the classifier decreases with the right selection of features. Furthermore, [6] studies are moving forward for prediction of epileptic seizure which is considered a more challenging problem.  Seizure prediction is being done on humans as well as animals. Analysis was done on feature extraction methods with linear and nonlinear under different stages of the epileptic seizure. Multiscale principle analysis and EEG decomposition [11] methods are used along with feature extraction and selection methods which require statistical features as well as empirical mode decomposition is used in decomposing the EEG signals. Various classifiers are used for the classification process of the extracted features. According to [16] review article, many automated epileptic seizure detection techniques are covered in time, frequency and time-frequency domain along with different classification techniques.

As this gruesome disorder prevails, it is important to detect the epileptic seizure at the right time before the situation of the individual worsens. Therefore, this study provides an analysis of the feature extracting methodologies used for the detection of epileptic seizure along with the ML techniques that are widely used in the prediction of epileptic seizures are studied and the performance of these techniques is analysed on different datasets. Furthermore, the prime focus of this study is to identify the gaps and challenges and pinpointing opportunities and further advancements that lead the researchers to better opportunities in this area.

This paper is organised in sections. Section II provides us with research methodology where the research objectives, research scheme, research questions, screening and selection methods of the papers and research string is generated and described in detail for researchers to get the gist of the whole paper. Figure 1 delineates the research process model. Section II gives the data analysis which includes the search results in phases as well as discussion and assessment of the research questions in detail. Section IV provides with the discussion of this paper where a taxonomy is also proposed. Section V discusses the issues and challenges of that were faced and Section VI concludes this whole research.

## II. RESEARCH METHODOLOGY

This systematic literature review provides a full procedure to abet the collection and investigation of the relevant articles from the considered studies. Figure 1 represents how the SLR is carried out. 1) Research Methodology: i) Research objectives ii) Research strategy iii) Research questions and motivation 2) Selection and Screening of papers: i) Inclusion/Exclusion criteria ii) Data extraction analysis 3) Assessment of Research questions: i) Justification of research questions.

### *A. OBJECTIVES OF THIS RESEARCH (R-O)*

Following are the chief purpose of this research:

R-O1: Understanding brain signals during epileptic seizure and differentiating normal seizure from an epileptic seizure.

R-O2: An understanding of the performance of Machine Learning classifiers used for classification.

R-O3: Features extraction techniques used for the correct epileptic seizure prediction.

R-O4: Identifying gaps and challenges in the already published research.

### *B. RESEARCH QUESTIONS (RQ)*

In order to accomplish a good SLR, research questions were formed as the first step of this research. Moreover, an exhaustive research planning was needed to further move on with this paper. Selection and screening process was done in which the inclusion and exclusion criteria in this research was described to further narrow down the research. Furthermore, the generated research questions are assessed and discussed thoroughly.

**TABLE I.** **Research Questions and the motivation behind them**

| | Research Questions | Motivation |
|---|---|---|
| **RQ1** | What machine learning classifiers are used in majority of the research for the diagnostics of epileptic seizure? | For the understanding of the model requirements for the prediction of epileptic seizure |
| **RQ2** | What kind of feature extracting methods are being used and what kind of features are being extracted from the EEG signal? | To understand the performance of classification process based on the features extracted |
| **RQ3** | What are the gaps and challenges in the detection of epileptic seizures? | This question aims to identify strengths and limitations in pattern recognition techniques and performance of classifiers based on feature extraction on various datasets. |
| **RQ4** | What datasets are used for epileptic seizure detection? | To check the biasness of dataset |

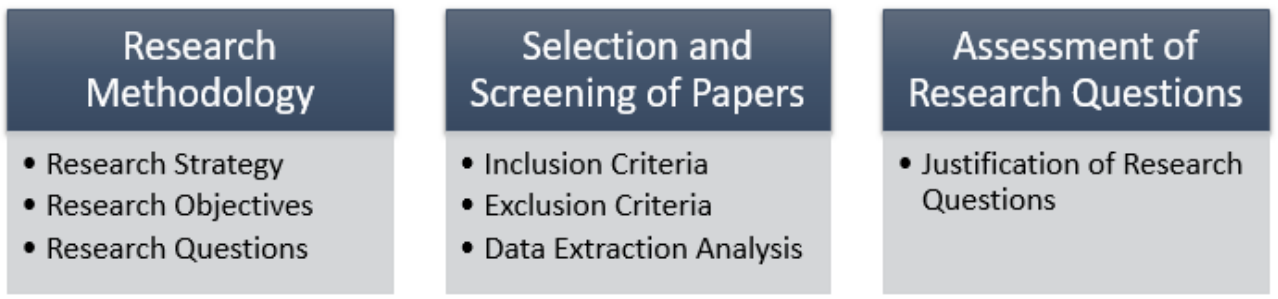

Figure 1 **SLR Process Representation**

### *C. SEARCH SCHEME*

The most significant part of a systematic literature review is to create a search strategy and executing that strategy in a systematic manner. Firstly, the goal is to collect as many relevant articles based on the chosen domain. The procedure further requires an illustration of search string, literature resources that are utilised for search application and the inclusion/exclusion criteria strategy to get the most significant and relevant articles from the pool of articles.

### 1) SEARCH STRING

A keyword based string was formulated in order to conduct an effective search to gather relevant studies by using five well known repositories. The analysis of chef concepts were done on the light of research questions to maintain authenticity. In Table II, keywords used for finding the relevant articles as well as their alternate words are described. The "+" sign portrays the inclusion criteria for studies that have said terms.

**TABLE II.** Terms and keywords used in search

| Terms (Keywords) | Synonyms / Alternate Keywords |
|---|---|
| **+ Machine Learning** | Classification Techniques, ML classifiers, Classification |
| **+ Epileptic Seizure** | Seizure detection, Epilepsy detection, Convulsions, Epileptic seizure detection |
| **+ Feature Extraction** | - |

To form a search string, the logical operators "AND" and "OR" were used as a combination of a finalized alternate keyword order to form a search string. The operator "OR" is the indication of additional options for the research and the "AND" operator is used for joining the terms in order to form the relevant search terms for relevant results.

The search string that is finalized contains four fragments .The first fragment is utilised to obtain the results that are related to machine learning and the second fragment connects with it and looks for results that include epileptic seizure, the third fragment includes results related to feature selection methods and the last fragment shows detection.

$$R = \forall\ [(ML\ \vee SL\ \vee DL) \wedge (ES\ \vee C\ \vee E)\ \wedge (FE\ \vee FS\ )\ \wedge (D\ \vee D\ \vee C)] \quad (1)$$

In the above equation (1), R represents the search results while '∀' represents 'for all', '∨' depicts 'OR' operator and '∧' for 'AND' operator and combining these search terms formulates the search string that is expressed in Table II. Generically, the search term in equation (1) can be expressed as:

((machine learning OR supervised learning OR deep learning) AND ("epileptic seizure" OR "convulsions" OR "epilepsy") AND ("feature extraction" OR "feature selection methods") AND ("diagnosis" OR "detection" OR "classification"))

### 2) LITERATURE RESOURCES

The journals that were selected for executing this research were from well-known journals and were selected from their online repositories. Their names and details are mentioned in Table II.

**TABLE III.** Publisher wise search strings

| *Data Repository* | *Relevant Search Strings* |
|---|---|
| *Science Direct* | *((MACHINE LEARNING OR SUPERVISED LEARNING OR DEEPLEARNING) AND ("EPILEPTIC SEIZURE" OR "CONVULSIONS" OR "EPILEPSY") AND (FEATURE EXTRACTION OR FEATURE SELECTION METHODS) AND (DIAGNOSIS OR DETECTION OR CLASSIFICATION))* |
| *Springer Link* | *((MACHINE LEARNING OR SUPERVISED LEARNING OR DEEPLEARNING) AND ("EPILEPTIC SEIZURE" OR "CONVULSIONS" OR "EPILEPSY") AND (FEATURE EXTRACTION OR FEATURE SELECTION METHODS) AND (DIAGNOSIS OR DETECTION OR CLASSIFICATION))FIELDS] [ALL FIELDS] )* |
| *IEEE Xplore* | *(((((("ALL METADATA":"MACHINE LEARNING") OR "ALL METADATA":SUPERVISED LEARNING ) OR "ALL METADATA":DEEP LEARNING) AND "ALL METADATA":EPILEPTIC SEIZURE) OR "ALL METADATA":CONVULSIONS) OR "ALL METADATA":EPILEPSY) AND "ALL METADATA":FEATURE EXTRACTION) OR "ALL METADATA":FEATURE SELECTION METHOD) AND "ALL METADATA":DIAGNOSIS) OR "ALL METADATA":DETECTION) OR "ALL METADATA":CLASSIFICATION)* |
| *MDPI* | *("MACHINE LEARNING"[ALL FIELDS] OR "SUPERVISED LEARNING" [ALL FIELDS] OR "DEEP LEARNING"[ALL FIELDS]) AND ("EPILEPTIC SEIZURES"[ALL FIELDS] OR "CONVULSIONS"[ALL FIELDS] OR "EPILEPY"[ALL FIELDS] OR "FEATURE EXTRCTION"[ALL FIELDS]) OR "FEATURE SELECTION METHODS" [ALL FIELDS]) AND ("DIAGNOSIS"[ALL FIELDS] OR "DETECTION"[ALL FIELDS] OR "CLASSIFICATION"[ALL FIELDS]* |
| *Wiley* | *((MACHINE LEARNING OR SUPERVISED LEARNING OR DEEPLEARNING) AND ("EPILEPTIC SEIZURE" OR "CONVULSIONS" OR "EPILEPSY") AND (FEATURE EXTRACTION OR FEATURE SELECTION METHODS) AND (DIAGNOSIS OR DETECTION OR CLASSIFICATION))* |
| *ACM Digital Library* | *(((("MACHINE" OR "SUPERVISED LEARNING") AND LEARNING) OR DEEP LEARNING OR (("MACHINE" AND "LEARNING") AND ((("EPILEPTIC SEIZURE" OR "CONVULSIONS") AND ("FEATURE " OR "FEATURE EXTRACTION") AND SELECTION) OR (("FEATURE EXTRACTION METHOD") AND CLASSIFICATION) OR ("DETECTION OF" AND ("CLASSIFICATION" OR "DIAGNOSIS"))))* |

### 3) INCLUSION AND EXCLUSION CREITERIA

Parameters defined for inclusion criteria (IC) are:

IC 1) Include studies that were primarily conducted for epileptic seizure prediction using machine learning techniques.

IC 2) Feature extraction methods targeting wavelet transform methods for the decomposition of EGG signals.

IC 3) Studies that encompasses machine learning classifiers that included RF, SVM, ANN and KNN.

The exclusion criteria applied on all the articles to exclude such EC 1) If the study did not involve any feature extraction techniques that involves wavelet transform techniques.

EC 2) The studies did not involve he classifiers such as RF, SVM, ANN and KNN.

### D. SELECTION OF RELEVANT PAPERS

In order to stay relevant with the topic, most research from the year 2017 to 2021 were selected for this paper. Various articles were selected based on the title and relevance with the said topic. As the process proceeded, the duplicate papers and some papers that did not meet the criteria of research were eliminated from the pool of papers. The papers that the criteria of research questions were thoroughly assessed and re-assessed and screening was done in order to acquire the relevant papers.

Firstly, the studies were filtered based on the titles and duplicate paper were removed. There were a bunch of irrelevant papers that were not related to the domain. In the selection process, a careful review of abstract was given and articles were included that described the prediction methods of epileptic seizure along with the feature extraction and classification techniques. Furthermore, the inclusion and exclusion criteria was defined which helped refining the articles.

### E. ABSTRACT BASED KEYWORDING

In abstract based keywording method, the abstract foregoes a thorough analysis in order to perceive the main idea of the article as well as the contribution it makes to that serves the domain and also to discover the most relevant keywords. Further, the keywords that are identified from the articles that were read previously and then combined due to the understanding of the research contribution in the domain. The main keywords that were chosen because of their direct importance with epileptic seizures were, (ML), (EEG), "Feature extraction" (FE) and "Automated Epilepsy Seizure Detection" (AESD).

## III. DATA ANALYSIS

In this section, the clear step by step evaluation of the selection criteria of articles is described in phases and it is described how the chosen articles were interrogated thoroughly in order to answer the research questions.

### A. SEARCH RESULTS

In the study of epileptic seizure detection, there have been steps that have been considered in order to carry out this research. Different well-known data sources were used, well-known datasets were collected and studied along with feature extraction and classification methods. The most important research component, which was the primary step of the research was that a total of 41 research papers were collected initially from different data sources online. The search was conducted in phases. In phase I, after the primary search part, title based selection was made where the title of the collected research papers were assessed from the 48 articles. In phase II, duplicate articles were removed and the articles that were irrelevant to the domain were also excluded based on the inclusion and exclusion criteria. In phase III, abstract based keywording was applied where the abstracts of the remaining articles were read and assessed thoroughly which left us with 36 papers altogether. Furthermore, in phase IV full text base analysis was done and the systematic literature review was conducted and proceeded on a total of 33 papers altogether. Figure 3 shows the digital library wise article selection chart which shows the articles were selected from the well-known repositories online.

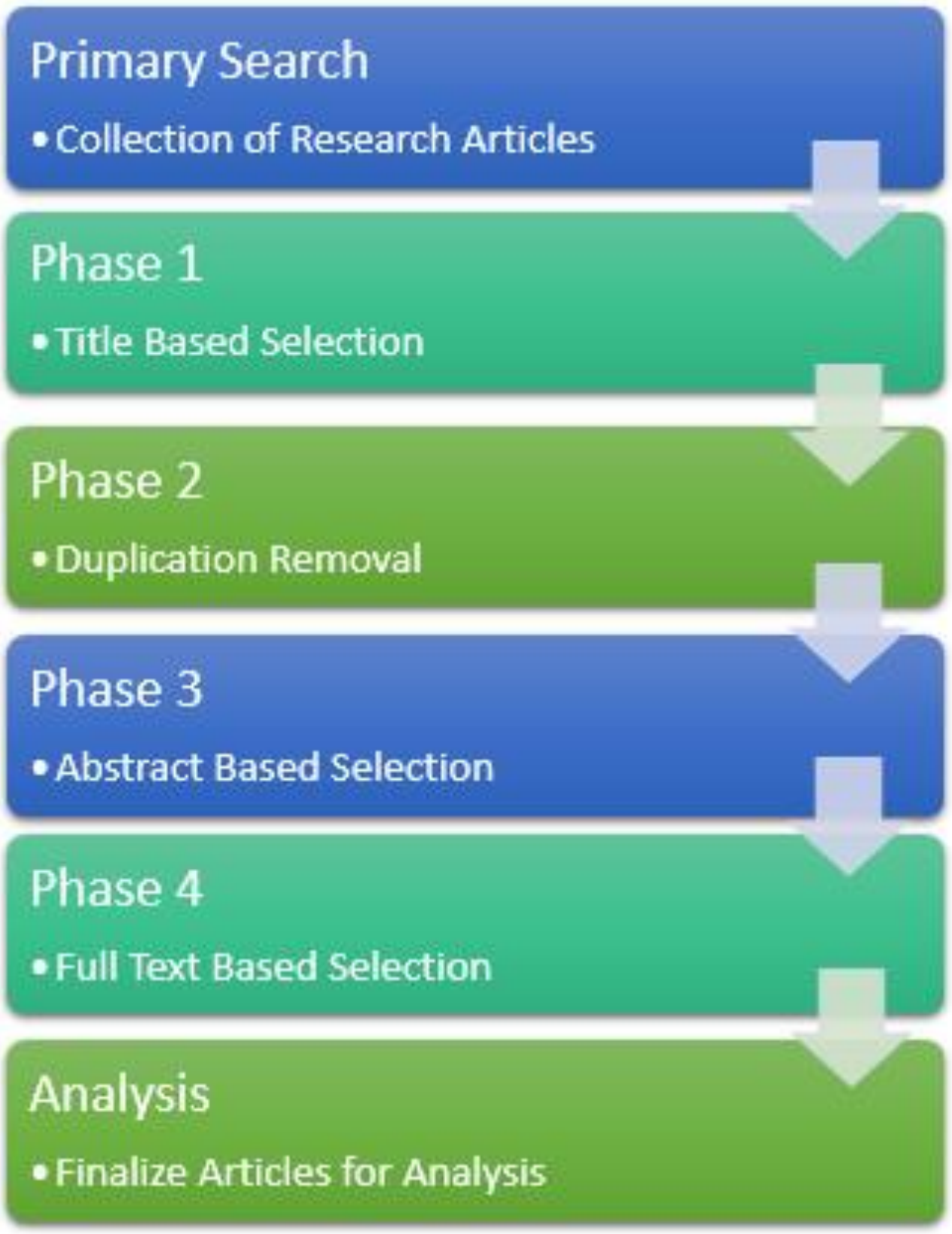

**Figure 2** Selection Procedure

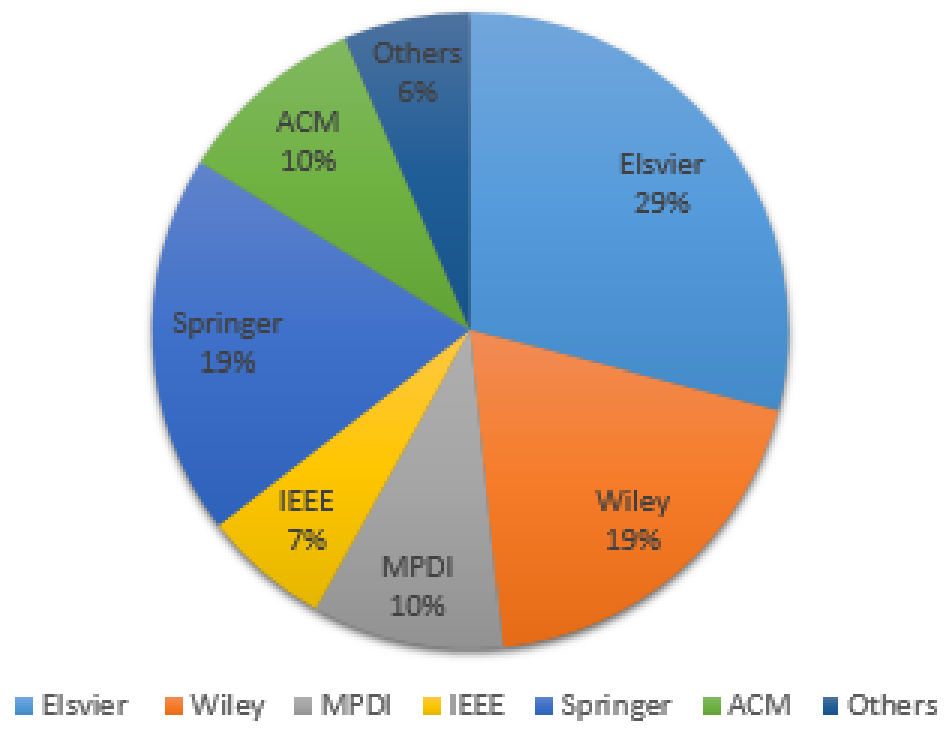

**Figure 3** Selected studies repository ratio

## IV. Assessment of Research Questions

In this section, the selected 33 articles were analyzed and assessed to justify the research questions. The facts obtained were later discussed in the assessment section of the research questions.

### 1) ASSESSMENT OF QUESTION 1. WHAT MACHINE LEARNING CLASSIFIERS ARE USED IN MAJOIRTY OF THE RESEARCH FOR DIAGNOSIS OF EPILEPTIC SEIZURE?

In this section, the machine learning classifiers are elaborated that are used in the detection of epileptic seizures. Classifiers The classifiers that have been shown extensively studied in the literature are shown in the table V along with the references.

#### *a: RANDOM FOREST (RF)*

This study is focusing on the performance of the selected classifiers in the detection process of epileptic seizures.

In [28], wavelet packet features technique is used and adopted random forest classifier as the epilepsy state classifier. They adopted a feature based splitting method for the tree nodes as well as generated a decision tree for each dataset. Splitting features were selected based on the criteria of gini gain. Classification accuracy obtained by random forest was 85% based on the number of decision trees provided which ranged from 50 to 1200. However, this classification was acquired in the pre-ictal stage while ictal and inter-ictal stages achieved 97% and 98% respectively and evidently performed better than other classifiers. Moreover, [31] used a MSC approach which is called multistage state classifier based on random forest algorithm. The structure of MSC includes three random forest classifiers that contain the basic logical decision thresholds which control the internal state transitions. The study followed cross frequency coupling and continuous wavelet transform for feature extraction and later used MSC model for classification. The MSC model was trained and optimization of two parameters was done through multi-iteration and 5-fold ROC cross validation over the training set. First in each ensemble random forest, the optimal number of estimators were found and later the optimal value for threshold was set. The testing of the model was done on Interictal time of 10 minutes and ictal time period of 66 seconds. The performance of the model was accessed based upon two groups, one which gained training and one with no training. It was stated that their model performed well with the first group with overall accuracy as 95%, group 2 achieved 79% accuracy and all patient sets achieved an overall 82% accuracy by using random forest algorithm based model for epileptic seizure detection.

Since Random Forest [9] is considered as a robust technique in selecting large features, therefore, it is widely used not only for classification but also for feature extracting methods for the better analysis of EEG signals in order to detect epileptic seizures. Random forest was used in this article for the selection of the top features in which out of 178 features, 20 variables were chosen and they were forwarded to ANN for further classification.

#### *b: SUPPORT VECTOR MACHINE (SVM)*

Support vector machine (SVM) is a technique in Machine Learning which is used widely and performs excellently in the areas where classification, prediction, estimation, regression and forecasting is involved [10]. A study [33] proposed SDI method and used SVM for classification. They used a radial basis kernel function (RBF) and determined hyperparameters by using the optimization technique. The class labels 0 and 1 were used to train the classifier for normal and epileptic seizure in EEG signal respectively. They provided training of each database separately by utilizing the method of cross validation and leave-one-subject-out method. A comparison was made with other linear approaches such as linear SVM, linear regression and linear discrimination. It was noted that SVM with (RBF) performed excellently than the compared classifiers giving sensitivity as 97.3%, false detection rate 0.4/h, F score of 97.22% and median detection delay was of 1.5s. The authors of [13] proposed a technique where they utilized empirical wavelet transform (EWT) method and decomposed the signal into rhythms and further used fast fourier transform (FFT) in order to determine the signal's frequency components. Wavelet functions are scaled at each segment and then the sub-band signals are reconstructed by using the EEG rhythms. The least square support vector machine is used in this study where the signal of EEG is classified into focal and non-focal signals. 50 pairs of focal and non-focal EEG signals were used here. They also used the same method on 750 signal pairs which lead to a classification accuracy of 82.53%, sensitivity of 81.60% and specificity of 83.46%. In [14] a hybrid SVM model is used in which the SVM has kernel type parameters and regularization constant C which leaves an impact on the performance. Parameter values are either default values or they are the values that are selected manually through trial and error. GA-SVM and PSO-SVM algorithms are used for the selection of these methods in this study for better classification techniques in epileptic seizure detection problem. However, hybrid SVM classifier, due to longer time for parameter selection, could not outperform GA-SVM and PSO-SVM where PSO-SVM performed slightly better than the rest achieving an accuracy of 99.38% and GA-SVM achieved 98.75% while SVM achieved 97.87% accuracy.

#### *c: k- NEAREST NEIGHBOUR (KNN)*

K-NN classifier is based on learning by analogy. It searches for the pattern space neighbours that are closest to a given unknown sample. Closeness is defined in terms of distance. The unknown sample is assigned the most common class among its neighbors [12]. The k-NN classifier [17] is used for the detection of any abnormality from the EEG signal during the epileptic seizure after the features are extracted from different levels of decompositions. The average accuracy obtained by the k-NN classifier after extracting three features extracted from the wavelet decomposition with sy4 wavelet at level 5 was 97.50% for detection of epileptic seizure. A new technique [20] for the detection of epileptic seizures has also been developed in which they used statistical features which were obtained by the discrete wavelet transform and feature reduction techniques such as PCA and LDA for the classification of normal EEG signals with epileptic seizure EEG signals by using k-NN and Naïve Bayse classifiers. The proposed method have shown high accuracy of classification by using LDA method for feature reduction and k-NN method for Bonn University database.

### d: ARTIFICIAL NEURAL NETWORK (ANN)

The detection of epileptic seizure [18] by the use of EEG signals is discussed in order to achieve the right approach to obtain classification accuracy of normal and epileptic seizures. For that, this study have used the discrete wavelet transform (DWT) feature extraction as well as GA-ANN for the process of selecting features that are more effective along with intended results that have an enhanced accuracy measure which can be achieved in both two class, which is epileptic seizure and normal and three class classification which is epileptic seizure, normal and seizure free. The 5-level decomposition method with db4 DWT was used also ANN and SVM contributed to provide accurate classification of 100% and 98.7% respectively with this method. A novel approach [22] is used for the EEG signal diagnosis for epileptic seizure by using the multi-DWT as well as genetic algorithm that is used with the four classification methods such as SVM, ANN, NB and KNN. The results in this study showed that ANN outperformed other classifiers with this technique. The EEG signals are firstly preprocessed which is the primary step of the method that helps in increasing the performance of the system and in removing the noises. The proposed system foregoes various stages for detecting epileptic seizure. Feature extraction is the second step which ends up generating a features matrix that later used in the classification process of EEG. The verification of success is done by the implementation of datasets in 14 combinations. The results were measured in terms of accuracy, specificity and sensitivity and it was noted that with this technique, ANN performed relatively better than the other classifiers. The accuracy achieved by ANN was 97.82% while SVM, NB and KNN achieved 97.15%, 97.32% and 97.58% respectively.

The authors of [23] trained the neural network algorithm and then tested it on 822 signals from the database. Five crucial features were used as input which were extracted from the EEG signals in time-frequency analysis and using continuous wavelet transform as well as subsequent statistical analysis. 583 which is 70% of the samples out of total samples were used for the system development and 239 which is 30% samples were used for testing of performance of the proposed model. The right parameters for ANN were selected by k-fold cross validation. Finally, the ANN classifier was implemented on the proposed model and according to the study, the results showed that high accuracy of 95.14% was achieved with ANN implemented on this model.

**TABLE IV. ML classifiers used in the selected articles**

| *Sr. No.* | *Classifiers* | *Used in* |
|---|---|---|
| **1** | SVM | [2], [3], [10], [13], [14], [15], [16], [18], [19], [21], [22], [26], [27], [29], [33] |
| **2** | RF | [8], [9], [10], [16], [19], [25], [28], [30], [31] |
| **3** | ANN | [9], [18], [22], [23], [26], [29] |
| **4** | KNN | [2], [3], [10], [12], [17], [19], [20], [21], [22], [26], [27] |

Domain importance was defined by seeing a gradual increase of interest in the said study field. Domain importance can be measured by gradually increased interest in the selected field of study. The published articles were selected in the years ranged from 2017 to 2022. Figure 5 portrays a year-wise frequency distribution chart of selected articles to the best of our knowledge, which is showing a non- linear increase and decrease of domain per year. The maximum number of papers out of 28 selected papers were identified in 2021, 5 papers included in the years 2017, 2018 and 2019. However, there is a gap seen in 2020 and 2022 where only 3 and 2 papers were identified respectively. Overall, it is shown that there is a rise and fall of the study in increasing years.

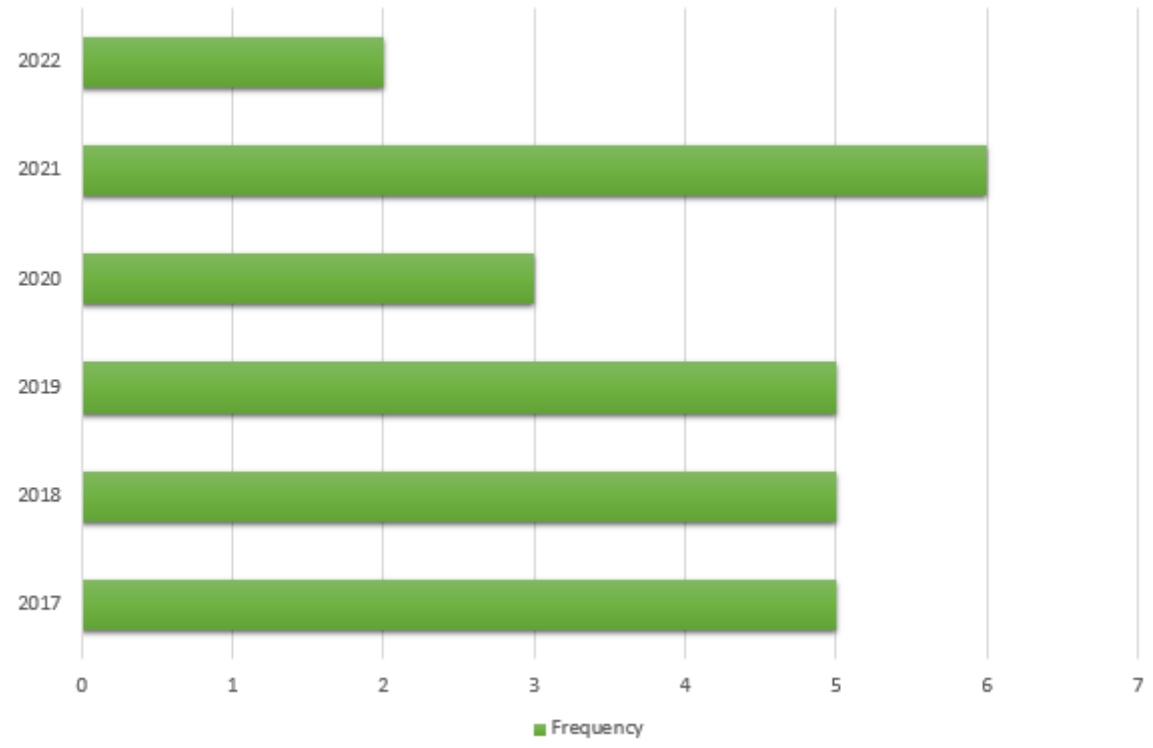

**Figure 4** A publication frequency of studies based on selected years

### 2) ASSESSMENT OF QUESTION 2. WHAT KIND OF FEATURES EXTRACTING METHODS ARE BEING USED AND WHAT FEATURES ARE BEING EXTRACTED FROM THE EEG SIGNAL?

This part of the paper will describe the features extraction techniques that are commonly used in literature.

A a multi-view deep feature extraction technique proposed for epileptic seizure detection [32] in real time. A model is introduced by the authors in which first the EEG signal is segmented by a fixed sliding window across the whole signal and passed two parameters which are window length l and window step s. It is stated that by increasing the length of the signal, the accuracy of the recognition process increases however it causes delay at real time applications. Therefore, they fixed the length of the sliding window l to 3s and step s to 1s. The study considered the time-frequency domain and used short-time fourier transform (STFT) to obtain spectrogram representation by converting the EEG signals through (STFT). For feature extraction, as soon as the spectrograms are generated, the proposed model can extract a set of deep features automatically by intra correlation and inter-correlation of the EEG channels which includes cross channel features and intra-channel features. For instance, given a single-channel EEG spectrogram, we can regard it as a spectral image where various spatial features are extracted within each fragment. In this step, high dimensional raw features are integrated into low dimensional latent characteristics with meaningful interpretation. An SSDA based channel selection methods is also introduced in order to filter the irrelevant features and to select channels that can extract critical features such as intra-channel features so the classification process can be more efficient. In this study, the accuracy obtained was 98.97% considering the used approach.

Furthermore, [13] empirical wavelet transform (EWT) is utilised and adaptive wavelets are constructed to extract different modes of the EEG signal. For instance, fourier transform (FFT) is used to extract the frequency components. Further, proper segmentation of the fourier segment is done in order to extract these modes. Moreover, the wavelet and scaling coefficients are gathered through their correspondence to each segment along with the reconstruction of sub-bands. The study primarily focused on focal and non-focal seizure by extracting the rhythm of the EEG signals using EWT. The features extracted in this study are non-linear thus it is the area that is generated using central tendency measure (CTM) generated by the 2D reconstructed phase space plot (RPS) giving an accuracy of 90%.

Moreover, this study utilized DWT with arithmetic coding [5] for feature extraction. They used db4 wavelet as an appropriate technique for non-stationary signals which is very effective in detecting sudden spikes that determine epileptic EEG signals. There were two types of statistical feature extraction methods were used in [4], that is, time and frequency domain statistical feature extraction and pattern adaptive wavelet transform feature extraction method. The methods of time and frequency domain are such that, the artifacts from the scalp EEG signal are removed by using the independent component analysis and the extracted features obtained were coefficient variance, mean and variance, root mean square, kurtosis, power, sum of mean, power spectral density and zero crossing rate. They considered 512 samples as the window size per rectangular window without the overlap with total 23 channels. Furthermore, the features that are extracted from these channels are used to form final features by averaging them with the same size of the window. The final features are further grouped into three signal frames that is normal, pre-ictal; that is about 30 seconds before the seizure starts and seizure frames throughout the training phase. The final features of the signal calculated during the testing phase are done just like how they were calculated in training phase. Moreover, in the pattern adapted wavelet based feature extraction method, a normal EEG signal based wavelet transform is constructed which has a length of 36 that is centered at 256 sample of the normal signal. Further, pattern adapted wavelet is applied to the signal in estimate the position and scale of the patter so that absolute values of wavelet coefficient can be formed. The duration of seizure is calculated as seizure duration = scale * sampling period and centered time = index * sampling period.

Therefore, a feature vector is constructed and seizures are detected along with root mean square, power spectral density and standard deviation being calculated as the extracted features.

A feature extracting strategy [19] is further proposed which is multimodal and used for the detection of epileptic seizures. We know how the complexity of EEG signals arise in terms of non-linear and non-stationary behaviour. This study have extracted the features which are situated on time domain, frequency domain, complexity based measures and wavelet entropy methods which are used in the classification process of healthy subjects and subjects that are foregoing an epileptic seizure heart rate oscillations that occur before seizure. The study also extracted non-linear features by using sample entropy which is based on KD tree algorithmic approach (fast sample entropy) as well as approximate entropy, and compared the performance with Wang et al (2017) and Hussain et al (2017b). Apparently, this method outperformed the results of Wang et al and were consistent with Hussain et al's result.

### 3) ASSESSMENT OF QUESTION 3. WHAT ARE THE GAPS AND CHALLENGES IN THE DETECTION OF EPILEPTIC SEIZURES?

This section gives you an analysis of the gaps and challenges faced in the study. The major gaps that have been witnessed, which can also be a bigger opportunity is that the techniques used majorly for the generalized epileptic seizure detection can be utilised to work solely on petit mal epilepsy, also known as absence seizures. To the best of our knowledge, limited data was observed on the detection of petit mal epilepsy. Usually, absence seizures have a 3 Hz per second spike wave discharge pattern [34] and often go unnoticed in kids because they occur for such a small duration of time, usually for 30 seconds. Hence, they are often missed by the guardians. However, they are frequent and may occur 10 to 30 times a day and is caused only in children, mostly females. It is sure that petit mal epilepsy begins at the time of childhood but sometimes it disappears before puberty. However, it is important to study and analyze the behaviour of children around the age of 4 to 14 because absence seizures can effect a child's daily life and it can become harder for them to cope with. The symptoms of this type of seizure includes the child staring blankly and having no idea of their environment.

Further challenges include that a large amount of dataset is required for validating the detection process of machine learning and deep learning techniques. However, to the best of our knowledge, that is not the case. Most datasets do not acquire a larger sample of EEG signals and contain signals in chunks and is deemed unsuitable for the real time EEG signal detection.

#### 4) ASSESSMENT OF QUESTION 4. WHAT DATASETS HAVE BEEN USED IN MAJORITY OF THE RSEARCH FOR EPILEPTIC SEIZURE DETECTION?

In [8], [5], [12], [25], [17], [21], [26], [18] considered the analysis of EEG signals on a publically available dataset from the University of Bonn. The dataset consists of five subsets labelled A, B. C. D and E. 100 single channel EEG segments are used in this dataset that contains a length of 23.6s and 4097 total samples per channel. Sets A and set B were the labels of EEG recordings of five healthy subjects with eyes open and eyes closed respectively. Datasets C and D contained patients who have epilepsy but are not foregoing a seizure currently. A,B and C were inter ictal stages The EEG recordings of C included the hippocampal formation from the hemisphere opposite to the epileptogenic zone and dataset D included the records of the epileptogenic zone meaning where the seizure usually arise. These recordings were collected at the time when the patients were not foregoing seizure, i.e. in seizure free time. Set E contains the collection of patients that are going through the seizure activity recorded by the hippocampal focus. These datasets were recorded by using 128 channels. Channels that consisted of pathological activities were removed from the computation. From the scalp EEG sets A and B the eye movement artifacts were also removed. The data was acquired by utilizing a 12-bit analog to digital converts with a 173.61 Hz sampling frequency. Further, they applied a bandpass filter on raw EEG data. EEG data consisted of 5 classes x 100 observations per class x 4097 (23.6 sec per observation). However, [29] utilised the same dataset but with spectral bandwidth that ranged from 0.5 Hz to 0.85 Hz of the EEG acquisition system and the sampling of data points was done at 173.61 Hz and passed to a 40Hz low pass filter. Dataset consisted of text files which were classified in binary as 0 or 1 meaning non-seizure and epileptic seizure respectively. With this data, two matrices were generated containing the sampled signal data Dataset from Children's Hospital called CHB-MIT EEG [4], [16], [24], [30], [32] scalp dataset was used in this study for experimentation. They recorded 22 patients at various times and made 654 files with uncontrollable seizures mainly that cannot be controlled by medicines. The files consists of recordings that start from 1 hour to 4 hours with 16 bit resolution and 256 sampling rate. This study included 1 female patients and 5 male patients that were aged from 1.5 to 22 years old. They used a standard international 10-20 system for the placement of sensors on the scalp and used that with 23 channels for the recording of the data. They carried out the experiment with a 23/256 duration of signal samples in each file. The first out of three datasets [15] where the EEG recordings were collected were from the Institute of Neuroscience, India. It included 19 channel scalp EEG which include a duration of 58 h taken from 115 patients which included 67 male and 48 females whose age ranged from 2.5 to 75 years old. Out of 115 subjects, there were 38 subjects suffering from epilepsy and 77 were healthy. The international 10-20 system of electrode placement along with the sampling rate of 128 Hz was used to collect the EEG signals. The second database was from CHB-MIT from the Physionet repository which is a publically available dataset. This contained 23 patients whose recording was made at a sampling rate of 256 Hz which included 844 h of data. They international 10-20 system bipolar montage was used on this dataset. The final database used in this study was obtained from TUH EEG which consisted of 316 patients following the same international 10-20 electrode placement system with 250 Hz sampling rate. From the EEG recordings, 222 out of 316 epileptic seizures were considered.

The epileptic EEG data [27] was being recorded of 16 subjects in Katip Celebi University, School of management and neurology department by using surface electrodes. Neurofax device was used to record the EEG data from various channels with the 100 Hz as sampling frequency. Electrodes were placed with regards to the international 10-20 system. Maximum 2 epochs were used on each patient in this study where each epoch contained 10 channels for 1 minute were used with a total of 32 epochs.

Further, in [1] a long term EEG recording of 275 patients were used from a public database called the European Union funded database. Moreover, EPILEPSIAE database was used by the researchers which is a widespread electroencephalography database of epilepsy patients.

**TABLE V. Applied approaches by the reviewed studies**

| Ref# | Year | Problem Tackled | Classifiers | Technique | Findings | Datasets |
|---|---|---|---|---|---|---|
| [2] | 2019 | Misdiagnose in manual methods so procedure is automated | SVM<br>KNN<br>Deep Neural Networks | Feature Scaling<br>Loss Function | SVM = 94% accuracy<br>KNN = 74% accuracy | Bonn University |
| [3] | 2019 | Differentiating normal EEG signals with Epileptic seizure signals in ictal and inter-ictal stage | Linear and Non-linear ML techniques NB, KNN, MLP, SVM | CAD based diagnoses using DWT, Wavelet Decomposition, Feature computation and classification, Arithmetic Coding | Overall 100% accuracy | Bonn University |
| [4] | 2019 | Identifying pre-ictal and ictal state of EEG signals | Fuzzy classifier | Pattern adaptive wavelet transform | 96% accuracy | CHB-MIT |
| [8] | 2018 | Seizure prediction and detection | Random Forest | DWT with 5 level decomposition | High classification sensitivity was achieved by this method reaching 99.95% in comparison with other studies | Bonn University. Freiburg Hospital. |
| [9] | 2021 | Automated framework created for automated detection of epilepsy | ANN model RPROP+, RPROP-, SAG, SLR | Feature Selection Back Propagation | SLR=99%, SAG=97% in balanced classes and SLR=87%, SAG=89% in balanced classes | Preprocessed Epileptic seizure recognition on UCI repository. Balanced and Imbalanced classes. |
| [10] | 2017 | Automated onset prediction | SVM MLP KNN RF | Multiscale principle analyzing for de-noising. EMD DWT Wavelet packet decomposition Inter vs. inter-ictal | With Freiburg dataset and CHB-MIT dataset, high accuracy has been achieved indicating both well-known datasets worked well with used technique | Freiburg Hospital. CHB-MIT. |
| [14] | 2017 | Hybrid model for epileptic seizure prediction | SVM | PSO-based SVM GA-based SVM DWT with db4 | SVM= 97.87% accuracy GA-SVM=98.75% PSO-SVM=99.38% | Publically available dataset |
| [15] | 2019 | Automated seizure detection process with comparison of proposed method with existing methods | SVM | Successive Decomposition Index SDI. Wavelet energy | Sensitivity achieved is 97.53% with F-measure with 97.22% | Ramaiah College Hospital. CHB-MIT. Temple Unit |
| [16] | 2022 | Detecting seizure and noon-seizure events | SVM RF | Tunable Q-wavelet transform | RF-sensitivity=91.5% RF-accuracy=93% SVM-sensitivity=9=89.2% SVM-accuracy=90.4% | CHB-MIT |
| [18] | 2021 | Detecting epileptic seizure and defining right features | SVM ANN | DWT 5 level and statistical calculations. Statistical features were extracted | 2-class: ANN=100% accuracy, SVM=100% accuracy 3-class: SVM=98.7% accuracy ANN=98.7% accuracy | Bonn University |
| [19] | 2018 | Monitoring of brain activity during epileptic seizure and normal state | SVM KNN Decision trees | Time-frequency domain characteristics | SVM=98% accuracy KNN=94% accuracy | Bonn University |
| | | | | Non-linear wavelet based entropy | | |
| [21] | 2020 | Brain activity at different regions for timely and accurate detection of epileptic seizure | SVM<br>KNN | Feature engineering (FT)<br>Wavelet transform<br>Sequential forward floating selection | SVM=99%, 100%, 100% in time, frequency, time-frequency respectively<br>KNN=99.5%, 99%,99.5% respectively | Bonn University |
| [22] | 2020 | Classifying normal brain signals with epileptic seizure while increasing accuracy and reducing computational cost | ANN<br>KNN<br>NB<br>SVM | 54-DWT wavelets<br>Derived features minimization by using Genetic Algorithm to select relevant features | ANN achieved higher accuracy reaching 97.82% in comparison with the rest. | 14 classification combinations using Bonn University dataset |
| [23] | 2020 | Effective real time epilepsy diagnosis | Feed forward multi-layer neural network, MLP, ANN | Field programmable gate array solution (FPGA) | 95% accuracy | TUH-EEG corpus database |
| [24] | 2020 | Predicting a seizure in pre-ictal stage in terms of specificity and sensitivity | Deep learning techniques<br>SVM<br>CNN | FT<br>EMD<br>WT<br>Feature extraction and handcrafted feature extraction methods. | Sensitivity=92.7%<br>Specificity=90.8% | CHB-MIT |
| [25] | 2018 | Development of automated seizure detection | Random forest | Synthesizing generalized stockwell transform (GST), singular value decomposition (SVD) based feature extraction.<br>Changing n values for 4 cases to see if it effects the accuracies | Highest classification accuracies are 99.12%, 99.16%, 98.65%, 98.62% for four cases | Bonn University |
| [26] | 2020 | Checking performance in terms of accuracy to find relevant patterns related to different mental activities using feature extraction.<br>Extracting features based on spectrogram. | K-means<br>SVM<br>Multilayer perceptron | STFT used and window parameters are set to obtain good results.<br>K-means to extract features.<br>Descriptors:<br>Spectral peaks<br>Frequency<br>Time | Comparison was done with other works and was noted that SVM kernel gave a better performance. | Bonn University |
| [27] | 2020 | Combining 4 different approaches to decompose non-linear and non-stationary signals into finite number of oscillations (IMFs) | SVM<br>KNN<br>NB<br>Logistic Regression | EMD and its DWT derivatives and use them to generate EEG into oscillations called IMF | EEMD provided with better accuracies than EMD analysis.<br>EEMD provided with a robust feature extraction and results | Kahib-Celebi school of medicine |
| [28] | 2019 | Pre-ictal stage prediction multi-class classification | Random forest | Wavelet packet features<br>Wavelet packed decomposition | 84% accuracy | CHB-MIT |
| [29] | 2021 | Detection of epileptic seizure | ANN, SVM, NN, CNN | Wavelet transform<br>Singular value decomposition entropy<br>Petrosian fractal dimension<br>Higuchi fractal dimension | ANN outperformed other classifiers | Bonn University |
| [30] | 2020 | Reducing seizure frequency or prevention of epileptic seizure by early prediction | Random forest<br>Decision trees | DWT<br>Coefficient of variance with all sub-bands | 99.81% accuracy | CHB-MIT |
| [31] | 2017 | Pre-clinical seizure state | Random forest | Cross-frequency coupling<br>Multistate classifier | Sensitivity=87.9%<br>Specificity=82.4%<br>Area under ROC=93.4% | Toronto western hospital epilepsy monitoring unit |
| [33] | 2019 | Computationally efficient automated seizure detection | SVM | Successive decomposition Index | SDI = higher detection rate of epileptic seizure in terms of sensitivity,<br>db1=97.53%<br>db2=97.28%<br>db3=95.80%<br>false-detection rate,<br>db1=0.4/h<br>db2=0.57/h<br>db3=0.49/h<br>median detection delay,<br>db1=1.5s<br>db2=1.7s<br>db3=1.5s<br>f-measure<br>db1=97.22%<br>db2=96.29%<br>db3=94.70% | CHB-MIT<br>Ramaiah Medical College<br>Temple University Hospital |

## VI. DISCUSSIONS

This section provides with a discussion of the study conducted from epileptic seizure detection. The problem of epileptic seizure detection was addressed and the fact how it is disturbing the daily life of people [6]. It being a neurological disorder necessitates the early detection of the disorder in order to prevent great harm and sometimes death [3]. In this study, various feature extraction methods have been discussed along with classifiers and datasets that are being utilized in the detection of an epileptic seizure. To the best of our knowledge, it has been shown that SVM, RF, ANN and KNN are mainly observed being used in various studies along with feature extraction techniques used for better detection of EEG signals on various datasets. Furthermore, most features are studied in time, frequency and time-frequency domain [7], [9], [11] which give good results. Further, the datasets used in majority are described along with their descriptions. It has been observed that many results vary with the variation of datasets picked for this problem. For this study, many wavelet based and similar techniques were targeted that were used in signal decomposition of the brain signal in order to understand various states of epileptic seizure and their prevention methods in the reviewed studies. Major techniques used to the best of our knowledge are mentioned in section III in complete detail. Moreover, a hierarchal representation of the conducted study is shown in figure 5.

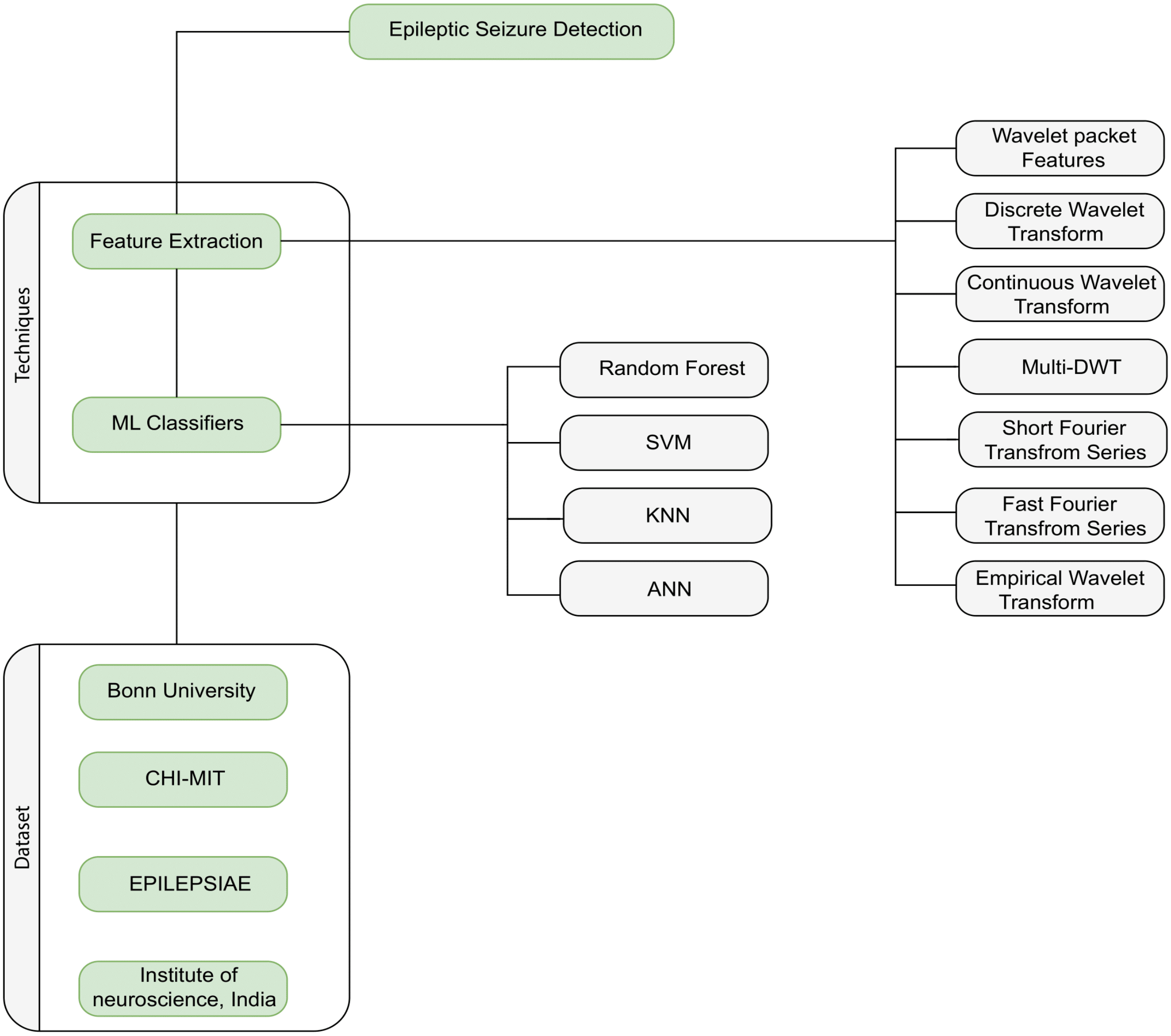

**Figure 5 a hierarchal representation**

## V. OPPERTUNITIES AND CHALLENGES

The opportunities lie for the researchers to get an idea is that they can study the prediction of epileptic seizure with the feature extraction techniques by studying the non-linear features more thoroughly and understand their results on different classifiers but mainly SVM and its other hybrid versions. Furthermore, a variety of datasets is used but due to different parameters set for feature extraction methods, it becomes difficult to gain insight on a larger dataset with a combination of feature extraction techniques. Also, this study was conducted to understand the basis of the how epileptic seizure detection methods are being used in the domain of machine learning. Further, for future work, researchers can work specifically on petit mal known as absence seizure and their detection that is somewhat a challenging task because of its minimal duration and negligible visual symptoms, however, the occurrence of these seizures is more frequent and can psychologically effect a child's life since it only occurs in kids from age 4 to 14. To the best of our knowledge, there hasn't been much literature regarding the ML techniques being applied solely for the detection of absence seizure. Major opportunity could be in the generation of a device which is user friendly for a home environment and not as terrifying as an electroencephalogram for children. Normally, children with absence seizures show mild signs or symptoms of absence seizures or abnormal behaviour such as staring blankly at the wall or not understanding a word someone says, a device should be made after testing the algorithms for the detection process with less chance of false positives so that the absence seizures can be detected and monitored at home by parents or guardians.

## VII. CONCLUSION

This study conducted a systematic literature review on detection of epileptic seizures which provides an analysis of the papers selected for this research in the field of epileptic seizure detection methods. The study was carried in a structured manner. An analysis of Ml classifiers was done along with the feature extraction methods being used in the study and the data source was thoroughly mentioned in the paper. Different datasets that are publically available were seen and how most of the studies selected have used these datasets in their research. Feature extraction techniques were mainly focused on techniques that used wavelet transform and signal decomposition was done for the prediction of epileptic seizure. The classifiers studied were SVM, RF, KNN and ANN and the major motivation was comparative studies which showed good results in with the feature extraction methods being used. Furthermore, it is suggested to study the most relevant predictive models for future so the direction for a quality research can be defined along with a suggestion to the future researchers to work on absence epilepsy in children and generating a separate dataset for this type of epileptic seizure.

## REFERENCES

[1] Sharmila, A. and Geethanjali, P., 2019. A review on the pattern detection methods for epilepsy seizure detection from EEG signals. *Biomedical Engineering/Biomedizinische Technik*, *64*(5), pp.507-517.

[2] Thara, D.K., PremaSudha, B.G. and Xiong, F., 2019. Auto-detection of epileptic seizure events using deep neural network with different feature scaling techniques. *Pattern Recognition Letters*, *128*, pp.544-550.

[3] Amin, H.U., Yusoff, M.Z. and Ahmad, R.F., 2020. A novel approach based on wavelet analysis and arithmetic coding for automated detection and diagnosis of epileptic seizure in EEG signals using machine learning techniques. *Biomedical Signal Processing and Control*, *56*, p.101707.

[4] Harpale, V. and Bairagi, V., 2021. An adaptive method for feature selection and extraction for classification of epileptic EEG signal in significant states. *Journal of King Saud University-Computer and Information Sciences*, *33*(6), pp.668-676.

[5] Boonyakitanont, P., Lek-Uthai, A., Chomtho, K. and Songsiri, J., 2020. A review of feature extraction and performance evaluation in epileptic seizure detection using EEG. *Biomedical Signal Processing and Control*, *57*, p.101702.

[6] Acharya, U.R., Hagiwara, Y. and Adeli, H., 2018. Automated seizure prediction. *Epilepsy & Behavior*, *88*, pp.251-261.

[7] Sharmila, A. and Geethanjali, P.J.I.A., 2016. DWT based detection of epileptic seizure from EEG signals using naive Bayes and k-NN classifiers. *Ieee Access*, *4*, pp.7716-7727.

[8] Tzimourta, K.D., Tzallas, A.T., Giannakeas, N., Astrakas, L.G., Tsalikakis, D.G., Angelidis, P. and Tsipouras, M.G., 2019. A robust methodology for classification of epileptic seizures in EEG signals. *Health and Technology*, *9*(2), pp.135-142.

[9] Mursalin, M., Zhang, Y., Chen, Y. and Chawla, N.V., 2017. Automated epileptic seizure detection using improved correlation-based feature selection with random forest classifier. *Neurocomputing*, *241*, pp.204-214.

[10] Alickovic, E., Kevric, J. and Subasi, A., 2018. Performance evaluation of empirical mode decomposition, discrete wavelet transform, and wavelet packed decomposition for automated epileptic seizure detection and prediction. *Biomedical signal processing and control*, *39*, pp.94-102.

[11] Li, M., Chen, W. and Zhang, T., 2016. Automatic epilepsy detection using wavelet-based nonlinear analysis and optimized SVM. *Biocybernetics and biomedical engineering*, *36*(4), pp.708-718.

[12] Tzallas, A.T., Tsipouras, M.G. and Fotiadis, D.I., 2009. Epileptic seizure detection in EEGs using time–frequency analysis. *IEEE transactions on information technology in biomedicine*, *13*(5), pp.703-710.

[13] Bhattacharyya, A., Sharma, M., Pachori, R.B., Sircar, P. and Acharya, U.R., 2018. A novel approach for automated detection of focal EEG signals using empirical wavelet transform. *Neural Computing and Applications*, *29*(8), pp.47-57.

[14] Subasi, A., Kevric, J. and Abdullah Canbaz, M., 2019. Epileptic seizure detection using hybrid machine learning methods. *Neural Computing and Applications*, *31*(1), pp.317-325.

[15] Raghu, S., Sriraam, N., Vasudeva Rao, S., Hegde, A.S. and Kubben, P.L., 2020. Automated detection of epileptic seizures using successive decomposition index and support vector machine classifier in long-term EEG. *Neural Computing and Applications*, *32*(13), pp.8965-8984.

[16] Pattnaik, S., Rout, N. and Sabut, S., 2022. Machine learning approach for epileptic seizure detection using the tunable-Q wavelet transform based time–frequency features. *International Journal of Information Technology*, pp.1-11.

[17] Harender, B. and Sharma, R.K., 2017, May. DWT based epileptic seizure detection from EEG signal using k-NN classifier. In *2017 International Conference on Trends in Electronics and Informatics (ICEI)* (pp. 762-765). IEEE.

[18] Omidvar, M., Zahedi, A. and Bakhshi, H., 2021. EEG signal processing for epilepsy seizure detection using 5-level Db4 discrete wavelet transform, GA-based feature selection and ANN/SVM classifiers. *Journal of Ambient Intelligence and Humanized Computing*, *12*(11), pp.10395-10403.

[19] Hussain, L., 2018. Detecting epileptic seizure with different feature extracting strategies using robust machine learning classification techniques by applying advance parameter optimization approach. *Cognitive neurodynamics*, *12*(3), pp.271-294.

[20] Sharmila, A. and Mahalakshmi, P., 2017. Wavelet-based feature extraction for classification of epileptic seizure EEG signal. *Journal of medical engineering & technology*, *41*(8), pp.670-680.

[21] Savadkoohi, M., Oladunni, T. and Thompson, L., 2020. A machine learning approach to epileptic seizure prediction using Electroencephalogram (EEG) Signal. *Biocybernetics and Biomedical Engineering*, *40*(3), pp.1328-1341.

[22] Mardini, W., Yassein, M.M.B., Al-Rawashdeh, R., Aljawarneh, S., Khamayseh, Y. and Meqdadi, O., 2020. Enhanced detection of epileptic seizure using EEG signals in combination with machine learning classifiers. *IEEE Access*, *8*, pp.24046-24055.

[23] Sarić, R., Jokić, D., Beganović, N., Pokvić, L.G. and Badnjević, A., 2020. FPGA-based real-time epileptic seizure classification using Artificial Neural Network. *Biomedical Signal Processing and Control*, *62*, p.102106.

[24] Usman, S.M., Khalid, S. and Aslam, M.H., 2020. Epileptic seizures prediction using deep learning techniques. *Ieee Access*, *8*, pp.39998-40007.

[25] Zhang, T., Chen, W. and Li, M., 2018. Generalized Stockwell transform and SVD-based epileptic seizure detection in EEG using random forest. *Biocybernetics and Biomedical Engineering*, *38*(3), pp.519-534.

[26] Ramos-Aguilar, R., Olvera-López, J.A., Olmos-Pineda, I. and Sánchez-Urrieta, S., 2020. Feature extraction from EEG spectrograms for epileptic seizure detection. *Pattern Recognition Letters*, *133*, pp.202-209.

[27] Karabiber Cura, O., Kocaaslan Atli, S., Türe, H.S. and Akan, A., 2020. Epileptic seizure classifications using empirical mode decomposition and its derivative. *Biomedical engineering online*, *19*(1), pp.1-22.

[28] Wang, Y., Cao, J., Lai, X. and Hu, D., 2019, June. Epileptic state classification for seizure prediction with wavelet packet features and random forest. In *2019 Chinese Control And Decision Conference (CCDC)* (pp. 3983-3987). IEEE.

[29] Rabby, M.K.M., Islam, A.K., Belkasim, S. and Bikdash, M.U., 2021, April. Wavelet transform-based feature extraction approach for epileptic seizure classification. In *Proceedings of the 2021 ACM southeast conference* (pp. 164-169).

[30] Hu, Z., Han, C., Guo, F., Qin, Q., Li, S. and Qin, Y., 2020, October. Epileptic Seizure Prediction from the Scalp EEG Signals by using Random Forest Algorithm. In *2020 13th International Congress on Image and Signal Processing, BioMedical Engineering and Informatics (CISP-BMEI)* (pp. 669-674). IEEE.

[31] Jacobs, D., Hilton, T., Del Campo, M., Carlen, P.L. and Bardakjian, B.L., 2018. Classification of pre-clinical seizure states using scalp EEG cross-frequency coupling features. *IEEE Transactions on Biomedical Engineering*, *65*(11), pp.2440-2449.

[32] Yuan, Y., Xun, G., Jia, K. and Zhang, A., 2017, August. A multi-view deep learning method for epileptic seizure detection using short-time fourier transform. In *Proceedings of the 8th ACM International Conference on Bioinformatics, Computational Biology, and Health Informatics* (pp. 213-222).

[33] Raghu, S., Sriraam, N., Vasudeva Rao, S., Hegde, A.S. and Kubben, P.L., 2020. Automated detection of epileptic seizures using successive decomposition index and support vector machine classifier in long-term EEG. *Neural Computing and Applications*, *32*(13), pp.8965-8984.

[34] Moeller, F., LeVan, P., Muhle, H., Stephani, U., Dubeau, F., Siniatchkin, M., & Gotman, J. (2010). Absence seizures: individual patterns revealed by EEG-fMRI. *Epilepsia*, *51*(10), 2000-2010.